\documentclass[%
    reprint,
 amsmath,amssymb,amsfont,
 aps,
]{revtex4-2}

\usepackage[normalem]{ulem}
\usepackage{siunitx}
\usepackage{graphicx}
\usepackage{dcolumn}
\usepackage{microtype}
\usepackage{textcomp}
\usepackage{newtxtext, newtxmath}

\usepackage{bm}
\usepackage{graphicx} 
\usepackage[dvipsnames]{xcolor}
\usepackage{todonotes}
\usepackage{tabularx}
\usepackage{booktabs}
\usepackage{gensymb}

\usepackage{hyperref}
\hypersetup{
    colorlinks=true,
    linkcolor=blue,
    citecolor=red,
    filecolor=magenta,
    urlcolor=cyan,
}
\usepackage[capitalise]{cleveref}

\let\vec\boldsymbol

\begin{document}

\title{Nonlinear Breit-Wheeler pair production using polarized photons from inverse Compton scattering}

\newcommand{\HIJ}{Helmholtz Institute Jena, Fröbelstieg 3, 07743 Jena, Germany}
\newcommand{\GSI}{GSI Helmholtzzentrum für Schwerionenforschung GmbH, Planckstraße 1, 64291 Darmstadt, Germany}
\newcommand{\IOQ}{Institute of Optics and Quantum Electronics, Friedrich-Schiller-Universität, Max-Wien-Platz 1, 07743 Jena, Germany}

\newcommand{\GU}{Department of Physics, University of Gothenburg, SE-41296 Gothenburg, Sweden}

\author{Daniel Seipt}
\email{d.seipt@hi-jena.gsi.de}
\affiliation{\HIJ}
\affiliation{\GSI}
\affiliation{\IOQ}

\author{Mathias Samuelsson}
\email{mathias.samuelsson@physics.gu.se}
\affiliation{\GU}

\author{Tom Blackburn}
\email{tom.blackburn@physics.gu.se}
\affiliation{\GU}

\date{\today}

\begin{abstract}
Observing multiphoton electron-positron pair production (the nonlinear Breit-Wheeler process) requires high-energy $\gamma$ rays to interact with strong electromagnetic fields.
In order for these observations to be as precise as possible, the $\gamma$ rays would ideally be both mono-energetic and highly polarized.
Here we perform Monte Carlo simulations of an experimental configuration that accomplishes this in two stages.
First, a multi-GeV electron beam interacts with a moderately intense laser pulse to produce a bright, highly polarized beam of $\gamma$ rays by inverse Compton scattering.
Second, after removing the primary electrons, these $\gamma$ rays collide with another, more intense, laser pulse in order to produce pairs.
We show that it is possible to measure the $\gamma$-ray polarization dependence of the nonlinear Breit-Wheeler process in near-term experiments, using a 100-TW class laser and currently available electron beams.
Furthermore, it would also be possible to observe harmonic structure and the perturbative-to-nonperturbative transition if such a laser were colocated with a future linear collider.
\end{abstract}

\maketitle

\section{Introduction}
    
Producing electron-positron pairs from the collision of photons is one of the most elusive processes in QED. Breit and Wheeler, in their seminal paper~\cite{Breit:PR1934}, were the first to consider two high-energy photons producing a pair as the inverse process of Dirac annihilation \cite{dirac_annihilation_1930}. Realizing this pair production in the laboratory is extremely challenging because the centre-of-mass energy of the colliding photons has to exceed the rest mass energy of the pair, $2mc^2\approx 1$~MeV.
The scarcity of sources of high-energy photons with suitable flux led Breit and Wheeler to deem such an endeavour ``hopeless''~\cite{Breit:PR1934}.

Pair production in the collision of a single high-energy photon (a $\gamma$ ray) with an intense laser pulse, first studied in Refs.~\citenum{Reiss:JMP1962,Nikishov:JETP1964a,Narozhnyi:JETP1965,Nikishov:JETP1967}, is called the ``nonlinear'' Breit-Wheeler (NBW) process, because the energy threshold can be overcome by absorbing many photons from the laser.
This process becomes efficient if the quantum parameter $\chi=|e|\sqrt{k_\mu F^{\mu\nu} F_{\nu\lambda} k^\lambda}/m^3$ reaches unity.
Here $k^\mu$ is the photon four-momentum, $F^{\mu\nu}$ is the laser field strength tensor, $e$ and $m$ are the electron charge and mass, respectively.
For more details on the theory of QED in strong laser fields and additional literature we refer to the recent reviews~\cite{dipiazza.rmp.2012,gonoskov.rmp.2022,fedotov.pr.2023}.

High-intensity lasers are now mature enough to make laboratory experiments on the Breit-Wheeler process feasible.
The seminal SLAC E-144 experiment reported the production of $\mathcal O(100)$ electron-positron pairs in collisions of TW laser pulse with the 46.6~GeV SLAC electron beam \cite{Burke:PRL1997}.
In this experiment, the high-energy photons that created the pairs were generated in same laser pulse (via nonlinear Compton scattering) as the pairs themselves.
Separating the $\gamma$-ray and pair production stages, so as to obtain a pure light-by-light interaction, has been considered extensively in the recent literature~\cite{blackburn.pop.2018,hartin.prd.2019,eckey.pra.2022,golub.prd.2022} and is now part of planned experimental programs~\cite{LUXE-CDR,LUXE-TDR,salgado_towards_2021}.
In fact, a two-stage scenario for pair production was already planned at E-144 but was never realized \cite{e144prop}.
Laser-based concepts for the direct observation of the linear Breit-Wheeler process have been considered both theoretically \cite{pike_photon-photon_2014,ribeyre.pre.2016,he.commphys.2021,sugimoto.prl.2023} and experimentally \cite{kettle_laserplasma_2021,dbt_mods_00051347,watt_bounding_2024}.
Furthermore, an experimental signal of the linear Breit-Wheeler process was reported in heavy-ion collision experiments, where the pair production can be considered as being due to the collision of ``quasi-real'' photons from the ions' Lorentz-boosted Coulomb fields \cite{star_bw}.

\begin{figure}
    \centering
    \includegraphics[width=0.95\linewidth]{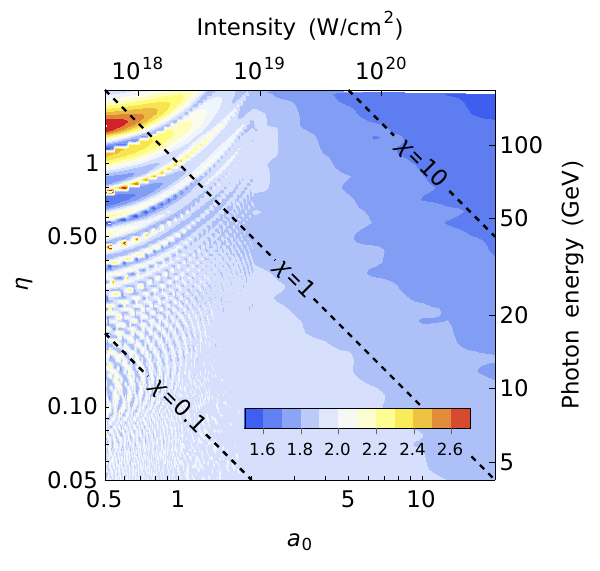}
    \caption{The ratio between $W_\perp$ and $W_\parallel$, the pair-creation probability rates for a photon with energy parameter $\eta = \chi / a_0$ that propagates through a linearly polarized plane electromagnetic wave with normalized amplitude $a_0$.
    The photon is either polarized perpendicular ($\perp$) or parallel ($\parallel$) to the electric field of the plane wave.
    The equivalent intensity and photon energy are calculated assuming a laser wavelength of 800~nm.}
    \label{fig:RateRatio}
    \end{figure}

A crucial aspect of the Breit-Wheeler process is the way it depends on the polarizations of the participating photons, as shown by Breit and Wheeler themselves~\cite{Breit:PR1934}.
If we consider pair production by a $\gamma$ ray colliding with a linearly polarized laser, the probability depends on the relative orientation of the $\gamma$-ray and laser polarizations.
The ratio of the pair-creation probability rates for a $\gamma$ ray that is polarized parallel or perpendicular to the laser electric field, $W_\parallel$ and $W_\perp$ respectively, is shown in \cref{fig:RateRatio}, where we introduce the laser normalized amplitude $a_0$ and an energy parameter $\eta = \chi / a_0$.
In the limit that $a_0 \gg 1$, the probability rates are related as $W_\perp = 2 W_\parallel$ for $\chi\ll1$ \cite{Reiss:JMP1962} and $W_\perp= \frac{3}{2} W_\parallel$ for $\chi\gg1$ \cite{Nikishov:JETP1964a}, showing that the probability is generally larger if the laser and $\gamma$-ray polarization are orthogonal.
Observing this difference is a goal of experiments aimed at precision measurements of QED in the strong-field regime~\cite{LUXE-CDR,LUXE-TDR} (see also Refs.~\citenum{gao_optimal_2022,zhao_signatures_2022,zhao_angle-dependent_2023}).

In this paper we discuss the prospects of measuring the polarization dependence of NBW in a two-stage configuration where the high-energy $\gamma$-ray beam is produced via \emph{linear} inverse Compton scattering (ICS) in a dedicated interaction point, separate from the strong-field focus. 
The $\gamma$ rays from such a source can be highly linearly polarized, parallel to the laser electric field, which permits easy control of the relative polarization orientation of the $\gamma$ rays and the second high-intensity laser. 
Using \emph{nonlinear} Compton scattering as a source of polarized $\gamma$ rays to study the polarization dependence of NBW has previously been considered in Refs.~\citenum{wan_high-energy_2020,qian_parametric_2023}.
Our study is motivated by the prospects of accelerator-based strong-field QED experiments, such as LUXE~\cite{LUXE-CDR,LUXE-TDR}.
The distinguishing features of our considered scenario are the $\gamma$ rays' polarization and their quasi-monoenergeticity. 
Throughout the paper we use natural units with $\hbar=c=1$. Scalar products between four-vectors are abbreviated as $k.p=k_\mu p^\mu$. We employ the Minkowski metric $g^{\mu\nu}=\text{diag} (1,-1,-1,-1)$.

\section{Proposed experimental configuration}

We consider here nonlinear Breit-Wheeler pair production, where a single photon with four-momentum $k$ interacts with $n$ photons of momentum $\kappa$ to produce an electron and positron with momenta $p$ and $p'$ respectively.
From the conservation of momentum, $k + n\kappa = p+p'$, one may show that a quantum energy parameter $\eta = k.\kappa/m^2 > 2/n$ is required in order to exceed the mass-energy threshold at given $n$ (neglecting for the moment intensity-dependent mass shifts).
In an asymmetric configuration where the $n$ photons are provided by a high-intensity optical laser, the single photon must have an energy of several GeV.
Furthermore, for efficient pair production one requires a quantum nonlinearity parameter $\chi = a_0 \eta \gtrsim 1$, at least for the multi-photon channels with $n \geq 2$.

    \begin{figure}[th!]
    \centering
    \includegraphics[width=1\linewidth]{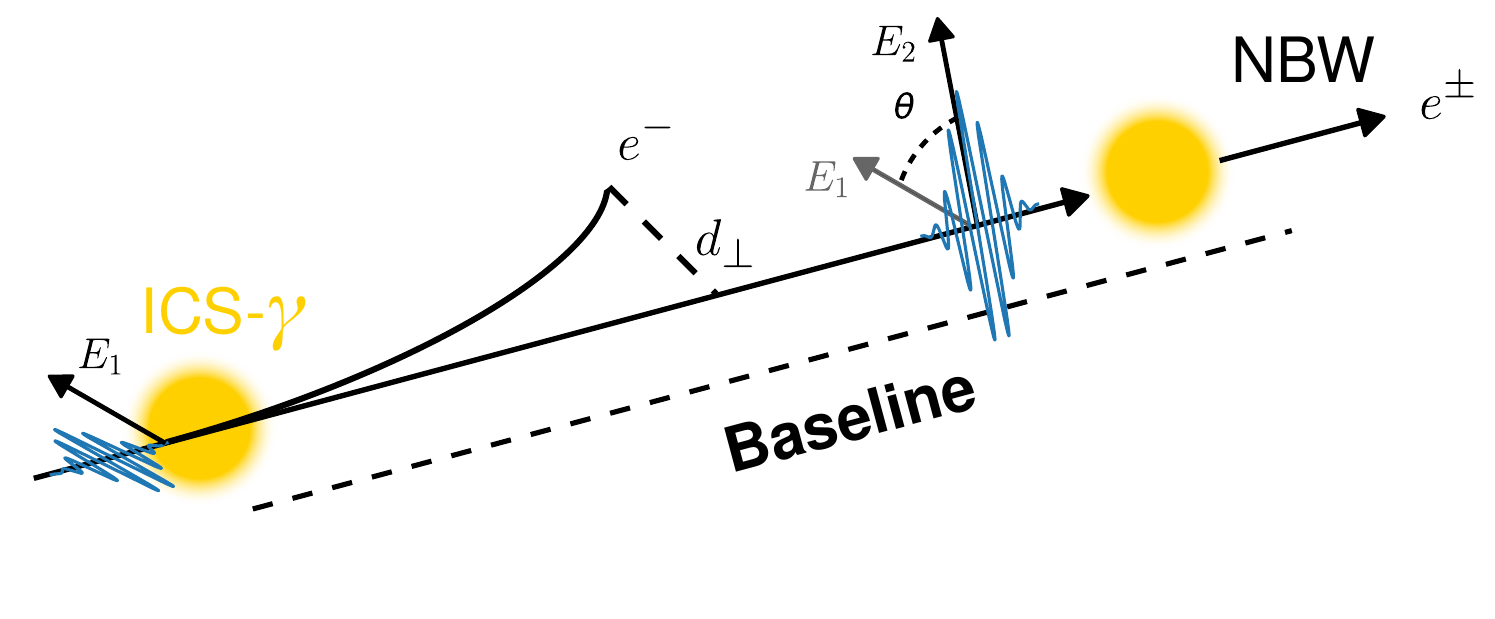}
    \caption{A laser pulse with polarization $E_1$ collides with an electron beam to create ICS $\gamma$ rays, which are predominantly polarised along $E_1$.
    After travelling along the baseline, the $\gamma$ rays collide with a second laser pulse, which is polarized along $E_2$ (at a pitch angle $\theta$ relative to $E_1$), and electron-positron pairs are created via the nonlinear Breit-Wheeler process (NBW).}
    \label{fig:schematic_setup_second_phase}
    \end{figure}

In addition to having multi-GeV energies, the high-energy photons ($\gamma$ rays) should ideally be bright, monoenergetic and highly polarized.
We consider therefore inverse Compton scattering off a multi-GeV electron beam, which fulfils all three, as the source (coherent bremsstrahlung is another possibility~\cite{gluex,borysov.prd.2022}).
The source is spatially separated from the strong-field interaction point (IP), where the $\gamma$ rays collide with a second laser pulse, by a baseline length $L$.
Electron and positrons are produced by the nonlinear Breit-Wheeler process at this second interaction point.
By rotating the polarizations of the two lasers, parameterized by the pitch angle $\theta$, we may alter the angle between the polarization of the $\gamma$ rays and the second laser.

A schematic of the experimental configuration is shown in \cref{fig:schematic_setup_second_phase}.
The baseline needs to be long enough that the primary electron beam, which creates the ICS-$\gamma$ rays, can be deflected away from the beam axis, while not so long that the flux of $\gamma$ rays at the second interaction point is too low for pair production to be observed.
An ultrarelativistic electron (Lorentz factor $\gamma \gg 1$) that travels a length $\Delta_B$ through a magnetic field of strength $B$ is deflected by an angle $\psi_B \simeq e B \Delta_B / (\gamma m)$.
The perpendicular distance between the electron beam and the original axis, $d_\perp$, assuming that beam travels freely for a distance $L$ after the magnetic field is
    \begin{equation}
    d_\perp\,\text{[m]} \simeq 0.3 \frac{B\,\text{[T]}~\Delta_B\,\text{[m]}}{E\,\text{[GeV]}} L\,\text{[m]}
    \end{equation}
where $B$ is the magnetic field strength and $E$ is the electron energy.
For $B \Delta_B \sim 1~\text{T}\text{m}$ and $E \sim 10~\text{GeV}$, we require baselines of several meters.

We describe the $\gamma$-ray and positron production stages in \cref{sec:PICA} and \cref{sec:Ptarmigan}, respectively.

\subsection{Gamma-ray generation stage}
\label{sec:PICA}

    \begin{table*}
    \caption{Collection of parameters for the three considered scenarios. This includes the properties of the primary electron-beam, the ICS laser, as well as the main characteristics of the produced $\gamma$ rays in stage 1.}
    \label{tab:simparams}
    \newcolumntype{C}{>{\hsize=0.4\hsize\centering\arraybackslash}X}%
    \newcolumntype{D}{>{\hsize=0.3\hsize\centering\arraybackslash}X}%
    \begin{tabularx}{\textwidth}{XDDCCC} \toprule
        & & & Case 1: LUXE     & Case 2: E144-like & Case 3: ILC-like \\ \midrule
        Electron energy        & $m\gamma$ & (GeV)                      & 16.5  & 50  & 200   \\ 
        Relative energy spread & $\Delta\gamma/\gamma$    &      & $10^{-3}$  & $10^{-3}$ & $10^{-3}$ \\ 
        Laser photon frequency & $\omega_\mathrm{ICS}$  & (eV)         & 4.1    & 1.55  & 1.55  \\ 
        Compton edge           & $\omega_\mathrm{max}$  & (GeV)  & 8.4  & 27  & 165   \\ 
        Laser spot size        & $w_{0,\rm ICS}$    & (\textmu{}m)             & 5  &  5  &5   \\ 
        Laser normalized vector potential        & $a_{\rm ICS}$    &              & 0.1  &  0.1  & 0.1   \\ 
        Laser FWHM duration    & $T$               & (ps)      & 1  &  1  & 1    \\ 
        Horizontal normzlized emittance   & $\varepsilon_x$  & (mm mrad)     & 1.4  & 1.4   & 0.03 \\ 
        Vertical normalized emittance     & $\varepsilon_y$  & (mm mrad)     & 1.4  & 1.4  & 10 \\ 
        Horizontal rms beam size   & $\sigma_x$   & (µm)          & 5  & 5  & 3  \\ 
        Vertical rms beam size     & $\sigma_y$   & (µm)          & 5  & 5   & 15  \\ 
        Beam length            & $\sigma_z$       & (µm)     & 20& 20   &  150  \\ 
        Beam charge            & $Q$             & (pC)      & 100   & 100  & 1000 \\ \midrule
        Baseline length        & $L$             & (m)      & 7.5  & 10  & 25  \\ \midrule
        Quantum energy parameter (at stage 2)          & $\eta$      &       & 0.1 & 0.32 & 1.95  \\
        Threshold harmonic (at stage 2, low $a_0$)     & $n_\star$    &                &  $\approx 20$ & $>6$ & $>1$  \\ \bottomrule
    \end{tabularx}
    \end{table*}

The purpose of the first stage is to generate a quasi-monoenergetic and linearly polarized GeV $\gamma$-ray beam via the inverse Compton scattering (ICS) process. Precise knowledge of the properties of the $\gamma$-rays at the strong-field IP is crucial for studying NBW in stage 2. Therefore we simulate the $\gamma$-ray properties using the Monte Carlo code PICA (Polarized ICS Calculator, available at Ref.~\citenum{pica}) taking into account realistic laser, electron beam and collision parameters. We simulate three different scenarios, with the simulation parameters collected in \cref{tab:simparams}.
    
In ICS, a low-energy photon from a laser with frequency $\omega_\mathrm{ICS}$ is scattered off a counter-propagating multi-GeV electron beam, which blueshifts the photon frequency to \cite{Corde:RevModPhys2013,Ranjan:PRSTAB2018}
    \begin{align} \label{eq:omegaICS}
    \omega(\vartheta) \simeq \frac{4\gamma^2\omega_\mathrm{ICS}}{1 + \gamma^2\vartheta^2 + 2\eta_\mathrm{ICS} + \frac{a_\mathrm{ICS}^2}{2}} \,.
    \end{align}
Here $\gamma=1/(1-\vec \beta^2)^{1/2}$ is the initial electron Lorentz factor and $\vartheta$ denotes the angle between the electron velocity $\vec \beta$ and the direction of the scattered photon.
Because the initial electron is ultra-relativistic, the scattered photons are beamed along $\vec \beta$ with a typical angular divergence $\vartheta \sim 1/\gamma \ll1$. The maximum photon energy is emitted in the electron forward direction $\vartheta=0$; this is called the Compton edge.
In Eq.~\eqref{eq:omegaICS} we also take into account the nonlinear intensity-dependent redshift~\cite{brown.pr.1964}, which becomes relevant as the ICS scattering laser normalized vector potential $a_\mathrm{ICS}$ approaches unity.
For our simulations this effect plays a negligible role, as we operate in the linear regime at $a_\mathrm{ICS}=0.1$.

The linear polarization of the ICS laser is (partially) transferred to the $\gamma$-rays. In the low-energy (Thomson) limit, characterized by $\eta_\mathrm{ICS}\ll1$, the emitted radiation has a dipole character. Thus for perfect back-scattering $\vartheta=0$, the $\gamma$ rays would be fully polarized.
Here $\eta_{\mathrm{ICS}}=p_0.\kappa_\mathrm{ICS}/m^2\simeq2\gamma\omega_\mathrm{ICS}/m$, the electron initial four-momentum $p_0=m\gamma(1,\vec \beta)$ and the laser four-momentum $\kappa_\mathrm{ICS}=\omega_\mathrm{ICS} (1,0,0,-1)$.
At high energy, $\eta_\mathrm{ICS}\sim 1$, it is known that the recoil of the photon imparted on the electron and spin-flip transitions lead to non-dipole contributions, reducing polarization degree.

The main objective for the ICS simulations is to accurately describe the photon flux, spectral bandwidth and polarization properties at the strong-field (stage 2) IP, which subtends a very small angular region, $\mathcal O(w_0/L) \ll 1/\gamma$, where $w_0$ is the focal spot radius of the strong field focus and $L$ is the baseline distance. To this end, it is necessary to accurately describe the initial electron beam energy distribution, transverse emittances in $x$ and $y$, as well as the longitudinal and transverse beam size at the ICS interaction point. A large part of the spectral bandwidth of the $\gamma$ rays stems from the electron beam emittance. For the parameters chosen for our simulation, see Table~\ref{tab:simparams}, the typical angular spread of the electrons in the beam $\sim \varepsilon_{x,y}/\gamma\sigma_{x,y}$ is comparable to the typical photon emission angle $\sim 1/\gamma$. Consequently, the value of $\vartheta$ for photons observed at a fixed global polar angle $\psi$ varies so much that $\omega(\vartheta)$ varies significantly over the electrons in the beam increasing the $\gamma$-ray bandwidth. In PICA the beam is sampled by a distribution of macro-electrons with weights chosen to match the total beam charge. 

For the laser pulse we employ the infinite Rayleigh approximation, which is a good approximation as long as $\omega_\mathrm{ICS} w_{0,\mathrm{ICS}}\gg 1$. Another crucial aspect for correctly predicting the $\gamma$ photon spectrum is to take into account the laser bandwidth due to the finite pulse duration $T$. In all our simulations the ICS laser is linearly polarized in the $x$-$z$ plane.
    
The PICA event generator creates macro-photons in a specified acceptance range ($\omega, \psi, \varphi$) using rejection sampling, where $\psi$ and $\varphi$ are the polar and azimuthal angles of the generated photon in the global reference frame. %
The event generation employs the Compton cross section for polarized incident photons, but the final state polarization is unresolved, as is the electron spin orientation \cite{book:Jauch,book:GreinerQED}.
For the polarization properties of the $\gamma$ rays we assign the Stokes parameters $\vec S$ for each macro-photon using the standard approach. In our convention the value of $S_1$ characterises the linear polarization component along 0$\degree$/90$\degree$ with respect to the $x$-$z$ plane, $S_2$ is for the 45$\degree$/135$\degree$ polarization component, and $S_3$ is for circular polarization. The macro-photons' weights, positions, momenta and Stokes parameters are written to HDF5-formatted output files for subsequent use in the pair production simulations, see \cref{sec:Ptarmigan}.

\subsection{Pair production stage}
\label{sec:Ptarmigan}

We simulate the pair production stage using the Monte Carlo particle-tracking code Ptarmigan (version 1.4.1, archived at Ref.~\citenum{ptarmigan}).
This code models the interaction of high-energy electrons, positrons and $\gamma$ rays with high-intensity laser pulses, taking into account the spatiotemporal structure of the laser pulse, the particles' classical dynamics and strong-field QED processes: for details see Ref.~\citenum{blackburn.pop.2023}.
In this work we use Ptarmigan in LMA (locally monochromatic approximation) mode, which includes interference effects at the scale of the laser wavelength and is therefore accurate in both the transition regime, $a_0 \simeq 1$, and the quasistatic regime, $a_0 \gg 1$~\cite{heinzl.pra.2020}.
In particular, this approximation allows us to resolve harmonic structure in the spectra of outgoing particles, which is a key signature of how pair production changes from two-photon, to multi-photon, and eventually tunnelling~\cite{blackburn.epjd.2022}.

Ptarmigan has recently been upgraded with the capability to import data for the individual particles that compose the incident beam, so that laser-collision simulations can be run using the output of other codes~\cite{ptarmigan}.
As discussed in \cref{sec:PICA}, PICA produces a Ptarmigan-compatible HDF5-formatted output file, which describes a beam of $\gamma$ rays in terms of their weights $w_i$, four-positions $x^\mu_i$, four-momenta $k^\mu_i$, and Stokes parameters $\vec{S}_i$.
Once the coordinate systems have been aligned, it is necessary only to specify the distance between the interaction points, i.e. the baseline length $L$, to run a simulation.
Ptarmigan imports the data, reconstructs the macroparticles, and then propagates them ballistically between the interaction points, up to the edge of the high-field region.
From there, the usual particle tracking and Monte Carlo sampling for strong-field QED processes takes place, continuing until the particles have left the high-field region~\cite{blackburn.pop.2023}.

This high-field region is created by a focused, linearly polarized laser with fixed energy $\mathcal{E}_0$.
The pulse has Gaussian spatial and temporal envelopes.
The laser spot size $w_0$ needed to reach a given $a_0$ may be determined as:
    \begin{align}
    w_0\,\text{[µm]}= 38.2 \, \frac{\lambda\,\text{[µm]}}{a_0} \sqrt{\frac{\mathcal{E}_0\,\text{[J]}}{\tau\,\text{[30 fs]}}}
    \label{eq:spotsize}
    \end{align}
where $\lambda$ is the laser wavelength and $\tau$ the FWHM pulse duration.
In our simulations, the laser intensity is varied in the range $0.5 \leq a_0 \leq 10$.
The pitch angle, the relative angle between the polarizations of the two lasers, is $\theta\in[0,\frac{\pi}{2}]$.

\section{Simulation results}

We simulated the polarization dependence of nonlinear Breit-Wheeler pair production for three different experimental scenarios, the parameters of which are summarized in \cref{tab:simparams}.
Our choices of parameters are guided by currently planned LUXE experiment \cite{LUXE-CDR,LUXE-TDR}, the seminal E-144 experiment at SLAC \cite{Burke:PRL1997}, and similar configurations that could become available at future linear colliders if one were to co-locate high-power lasers near them.

\subsection{Case 1 (LUXE)}
    
    \begin{figure*}[ht!]
        \centering
        \includegraphics[width=\textwidth]{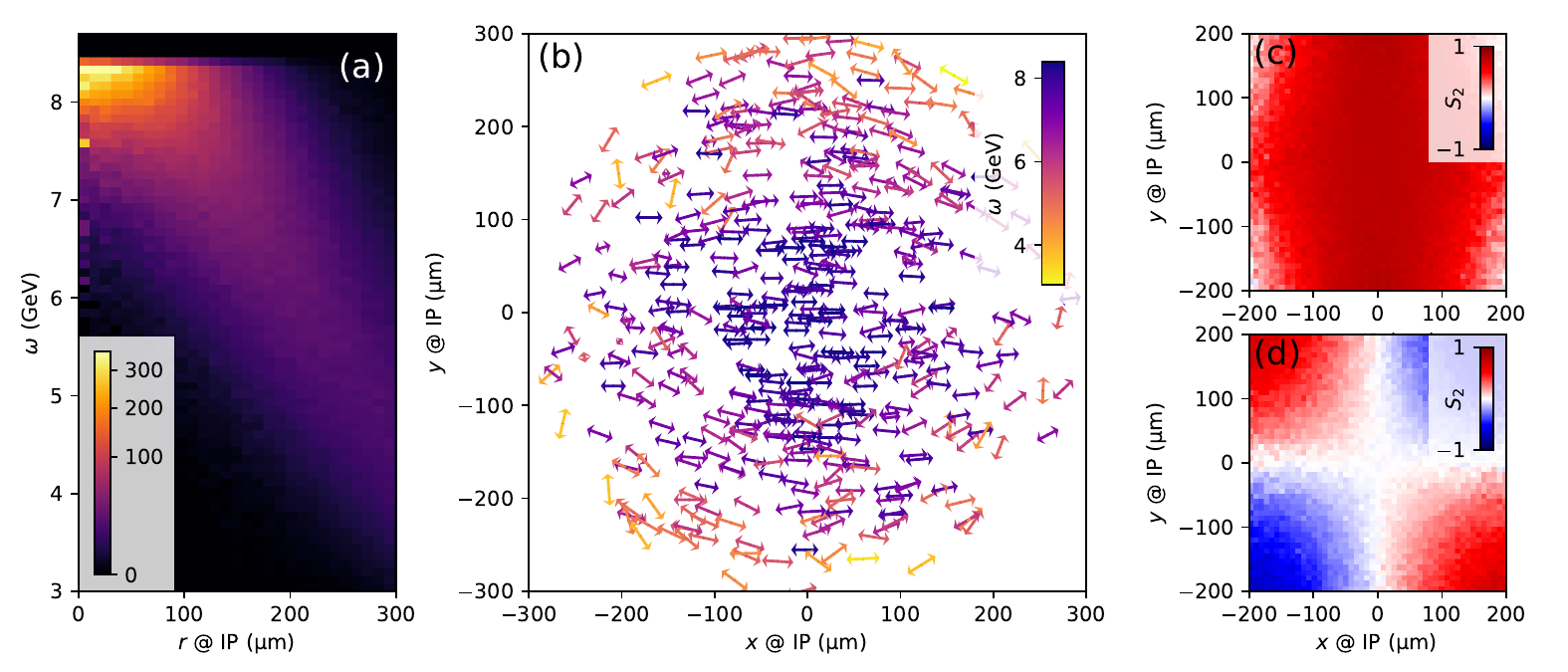}
        \hspace*{-0.2cm}\includegraphics[width=0.95\linewidth]{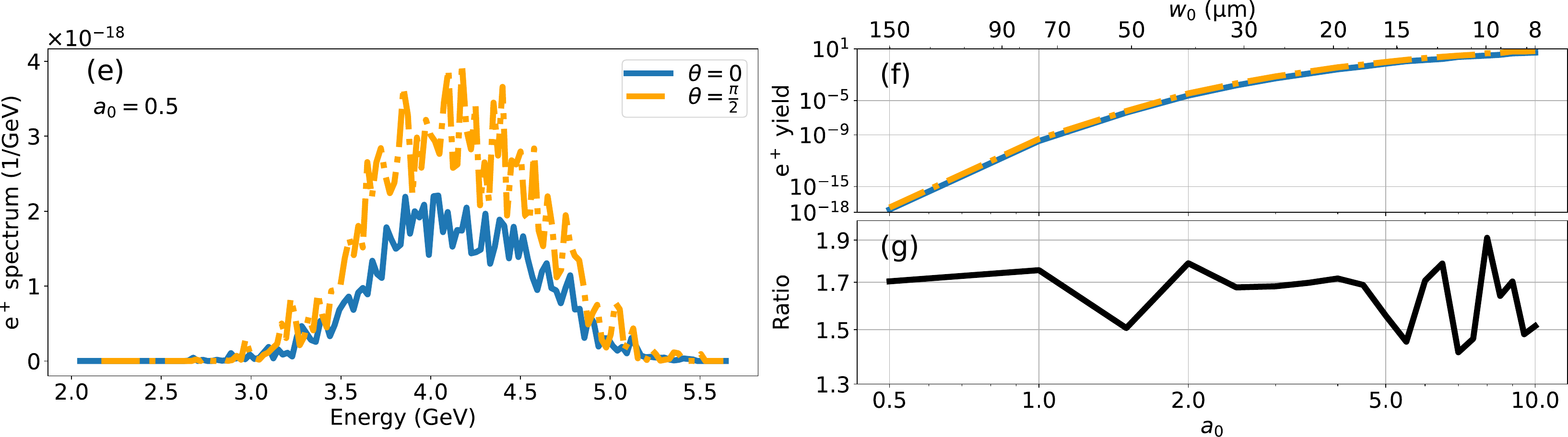}
        \caption{%
            Properties of the $\gamma$-ray beam at the strong-field IP (a-d) and of the positrons they produce (e-g) for case 1 (LUXE).
            (a) The radially resolved $\gamma$-ray energy spectrum.
            (b) Orientation and magnitude of the polarization (arrow orientation and length), as well as the energy (colour), of a subset of photons at the IP.
            (c) and (d) Stokes parameters $S_1$ and $S_2$, respectively.
            (e) Positron spectrum for $a_0=0.5$ and two pitch angles $\theta =0,\, \pi/2$.
            (f) Total positron yield as a function of $a_0$ (and laser spot size $w_0$) for the two pitch angles.
            (g) The ratio of the yields at $\theta = \pi/2$ and $\theta =0$.
        }
        \label{fig:case1}
    \end{figure*}

We begin by investigating a LUXE-like setup. The LUXE experiment has two main planned modes, $e$-laser and $\gamma$-laser, as well as an ICS-laser mode. In $e$-laser mode, high-energy electrons from European XFEL are directed to collide with a powerful laser, which produces high-energy photons and electron-positron pairs within the same laser field. In $\gamma$-laser mode,
the electron beam interacts with a thin foil to produce a broad band of bremsstrahlung photons. The electrons are diverted by a magnetic field and the photons collide with the laser pulse downstream to produce pairs \cite{LUXE-CDR, LUXE-TDR}.

To enable ICS-laser mode, a fraction of the main laser is extracted, frequency-tripled to 4.1 eV using a nonlinear crystal, and collided with the electron beam to produce high-energy photons via inverse Compton scattering. The advantage of a source like this is that it has a narrow energy distribution, high photon flux with a high degree of polarization \cite{LUXE-CDR}. The ability to rotate the polarization of the laser pulse allows us to control the $\gamma$-ray polarization.

Our simulation results for case 1 yield a mean photon energy of about 8 GeV at the center of the strong-field IP, with a root-mean-square (rms) bandwidth of approximately 0.5 GeV, see \cref{fig:case1}(a)-(b). The central part of the $\gamma$-ray beam has a photon fluence of $40$~photons/µm$^2$ per bunch crossing (BX). The photons are highly linearly polarized along the $x$-axis with a polarization degree 
    \begin{align}
    \bar{S} = \sqrt{\bar S_1^2 + \bar S_2^2} \approx 77~\%\,,
    \end{align}
where the bar denotes an average over all photons in a circular spot of radius 4 µm in the IP plane.

As can be seen in \cref{fig:case1}, photons with energies above 8 GeV can be found in quite a large spot at the IP. In the periphery of the strong-field focus we find photons of lower energy, mainly due to the angular dependence of the scattered photon frequency according to Eq.~\eqref{eq:omegaICS}. In this region the photons' Stokes vectors are aligned more in a radial pattern. When averaging over a 100-µm spot, the mean polarization degree is $\approx 75$~\%, which decreases to $\approx 60$~\% for a 300-µm spot. Panels (c) and (d) show two-dimensional histograms of the Stokes parameters $S_1$ and $S_2$ at the IP. While $S_1$ is nearly constant in the center with value $S_1\simeq0.77$, the values of $S_2$ are nearly zero in the center, and can be positive or negative in the periphery. This is in agreement with the observed polarization pattern in (b) since the polarization tilt angle (w.r.t. the $x$ axis) is given by $\frac{1}{2}\arctan (S_2/S_1)$. Naturally, the average polarization degree over a larger focal spot is significantly reduced due to the radial pattern (misalignment of the Stokes vectors). What this means for the pair production process is that for larger focal spots (smaller $a_0$) the effective polarization degree of the $\gamma$ rays is lower. However, this is in general quite complicated since the periphery has lower $\gamma$-ray energies and therefore the pair production rate can change significantly over the focus.

The positron yield at the second interaction point is simulated using the collision parameters given in the LUXE Conceptual Design Report~\cite{LUXE-CDR}, namely, a baseline length of 7.5 m and laser parameters corresponding to the ``phase 0'' configuration (peak power 40~TW, duration 30~fs).
The energy spectrum of the positrons produced when ICS-$\gamma$ rays collide with a laser of $a_0= 0.5$ is shown in \cref{fig:case1}(e).  Changing the pitch angle $\theta$ from 0 to $\pi/2$ increases the positron yield without changing the shape, which is mostly symmetric in both cases. However, there are no distinguishable harmonics here, because the lowest allowed harmonic order,
    \begin{align}
        n_{\star}=\left\lceil \frac{2\left(1+a_0^2/2\right)}{\eta} \right\rceil \label{eq:nharmonic},
    \end{align}
is $n_\star = 23$ for $a_0 = 0.5$ \cite{blackburn.epjd.2022}, which also explains the suppressed yield.

In \cref{fig:case1}(f), the total number of positrons is shown as a function of $a_0$. The laser spot size required to achieve a given $a_0$, according to \cref{eq:spotsize}, is shown on the upper horizontal axis. The positron yield increases exponentially as one would have expected from theory. At $a_0 = 0.5$, increasing the pitch angle from $0$ to $\pi/2$ increases the positron yield (per bunch crossing BX) from $0.2$ to $0.3$.
At $a_0 = 10.0$, the increase is from $3.8$ to $5.8$.
These yields are significantly lower than what is expected if the LUXE electron beam collides directly with the laser: at $a_0 = 5.0$, for example, the number of positrons is already $3.0 \times 10^3$~\cite{LUXE-TDR}. %
However, they are similar in magnitude to those expected in $\gamma$-laser mode, where the positron yield per BX is predicted to be $0.91$ at $a_0 = 5.0$, and $5.1$ at $a_0 = 10.0$~\cite{LUXE-TDR}. 

In \cref{fig:case1}(g) we show the ratio between the yields at $\theta =  0$ and $\pi/2$. The amplification is mostly constant over the entire range of $a_0$, with an average value of 1.7. This difference would be observable at LUXE, given the expected precision, statistics, and sustained operation that are planned.

\subsection{Case 2 (E144-like)}

    \begin{figure*}[ht!]
        \centering
        \includegraphics[width=\textwidth]{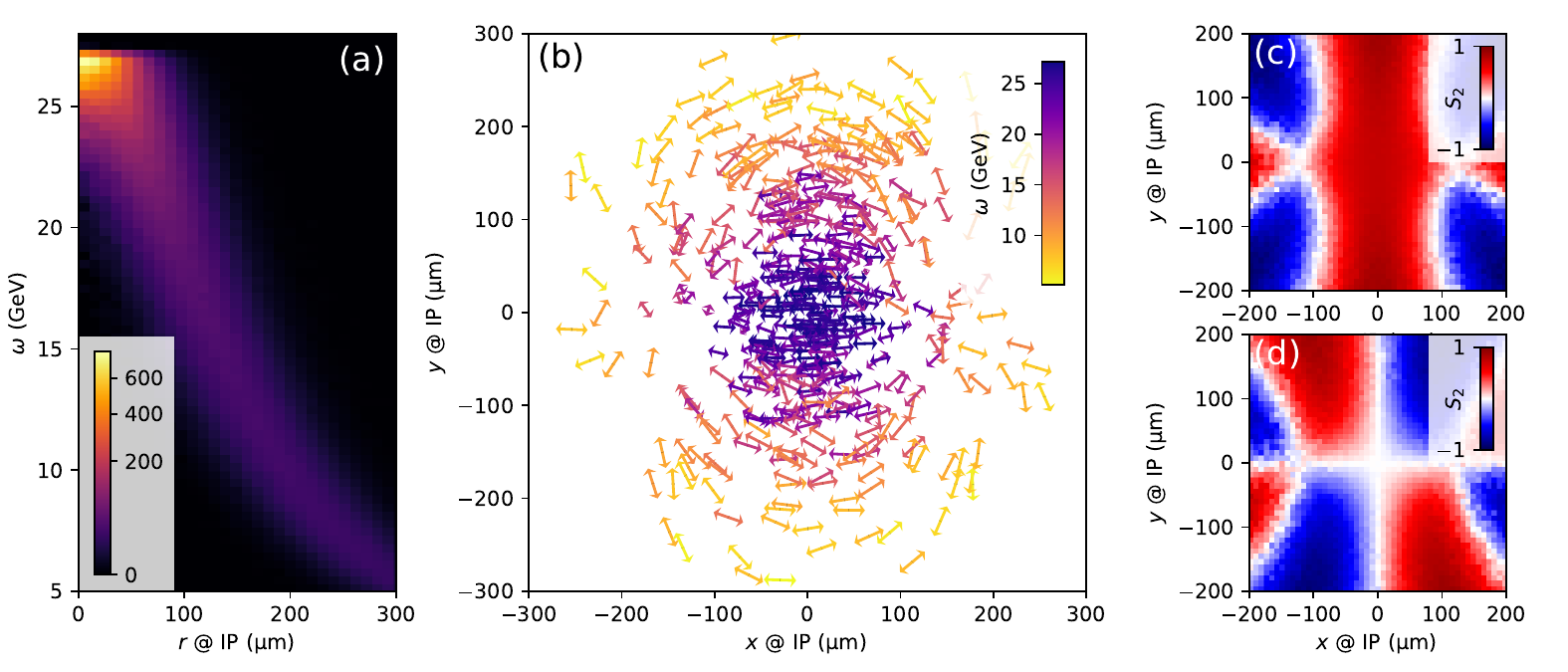}
        \hspace*{-0.45cm}\includegraphics[width=0.96\linewidth]{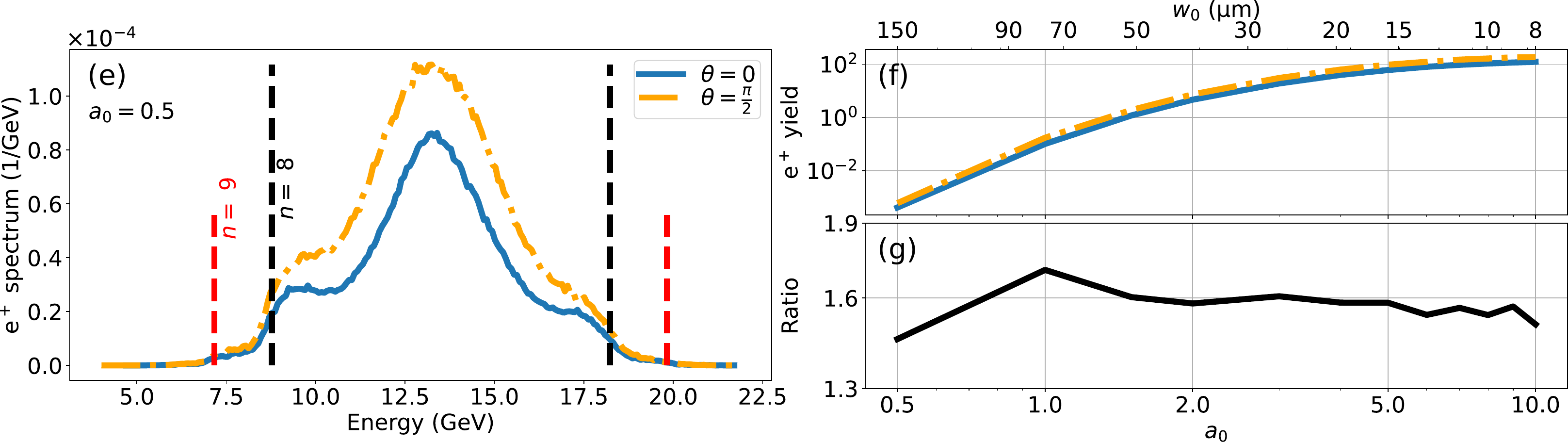}
        \caption{%
            Properties of the $\gamma$-ray beam at the strong-field IP (a-d) and of the positrons they produce (e-g) for case 2 (E144-like).
            (a) The radially resolved $\gamma$-ray energy spectrum.
            (b) Orientation and magnitude of the polarization (arrow orientation and length), as well as the energy (colour), of a subset of photons at the IP.
            (c) and (d) Stokes parameters $S_1$ and $S_2$, respectively.
            (e) Positron spectrum for $a_0=0.5$ and two pitch angles $\theta =0,\, \pi/2$.
            (f) Total positron yield as a function of $a_0$ (and laser spot size $w_0$) for the two pitch angles: vertical dashed lines give harmonic boundaries.
            (g) The ratio of the yields at $\theta = \pi/2$ and $\theta =0$.
        }
        \label{fig:case2}
    \end{figure*}

Here we consider consider a similar setup to case 1, but with an increased electron beam energy of 50~GeV, like the SLAC linac used in the E144 experiment \cite{Burke:PRL1997}. 
The main difference in this case is that the ICS laser does not have to be frequency tripled first before producing the $\gamma$ rays in the ICS stage: for all parameters see \cref{tab:simparams}. 

The $\gamma$ ray beam from the PICA simulation has a mean energy of 25.8~GeV in the center, with an rms energy spread of 1.18~GeV, and a photon flux of 69 photons$/$µm$^2$. In the central region the polarization degree is 73~\% and the photons are polarized dominantly along the $x$-axis, which is comparable in magnitude to case 1. For larger spots, the polarization degree rapidly decreases, falling to 64\% when averaging over a 100-µm spot, and only 37~\% for a 300-µm spot.
This stronger decrease compared to case 1 is due to the different polarization pattern, which can be seen in \cref{fig:case2}(b).
The main reason for this difference is the relatively smaller baseline length $L$ ,such that the ratio $L/\gamma=102$~µm compared to $L/\gamma=232$~µm for case 1. This brings the polarization nodes of the forward boosted dipole emission pattern closer to the center of the IP spot, which results in a stronger misalignment of the polarization axes of photons in the periphery of the IP focal plane.

In order to simulate the pair production stage, we assume a nominal baseline length of 10 m, but the laser parameters are kept identical to case~1. The main effect of larger baseline length $L$ is to reduce the positron yield by a factor of $L^{-2}$. In \cref{fig:case2}(e) we show the energy spectrum of the positrons for $a_0 = 0.5$. The yield is much higher than in case 1 because of the higher photon energy. There is now clearly visible harmonic structure, which is calculated to be of order $n = 8$. The energy interval spanned by this harmonic is given by
    \begin{align}
        \frac{1}{2}&\left[1-\left(1-4 / s_n\right)^{1 / 2}\right]<s<\frac{1}{2}\left[1+\left(1-4 / s_n\right)^{1 / 2}\right]\label{eq:Espan},
    \end{align}
where $s \approx {\epsilon_+}/{\omega}$, $\epsilon_+$ is the positron energy, $\omega$ is the photon energy at the Compton edge from \cref{tab:simparams}, $s_n= 2 n \eta / (1+a_0^2/2)$ and $n$ is the harmonic order as in \cref{eq:nharmonic}. These bounds are shown by vertical dashed lines in \cref{fig:case2}(e) with good agreement. Note that the positron spectrum is slightly asymmetric for both pitch angles, being higher at the lower bound. This is caused by the asymmetric energy spectrum of the photons at the interaction point: in \cref{fig:case2}(a) one can see that larger focal spots encompass more of the lower energy part of the spectrum.

In \cref{fig:case1}(f) the total number of positrons  is shown as a function of $a_0$, where we see that the total yield is much higher than the LUXE-like setup. The yield for $a_0 = 5.0$ is 60.6 and 95.9 positrons for pitch angle $\theta = 0$ and $\theta = \pi/2$ respectively, and for $a_0= 10.0$ the yield is 123.7 and 186.3 positrons. The ratio between the yields, shown in \cref{fig:case1}(g), is around 1.66, which is lower than case 1 because $\eta$ is larger. 

\subsection{Case 3 (ILC-like)}

    \begin{figure*}[ht!]
        \centering
        \includegraphics[width=\textwidth]{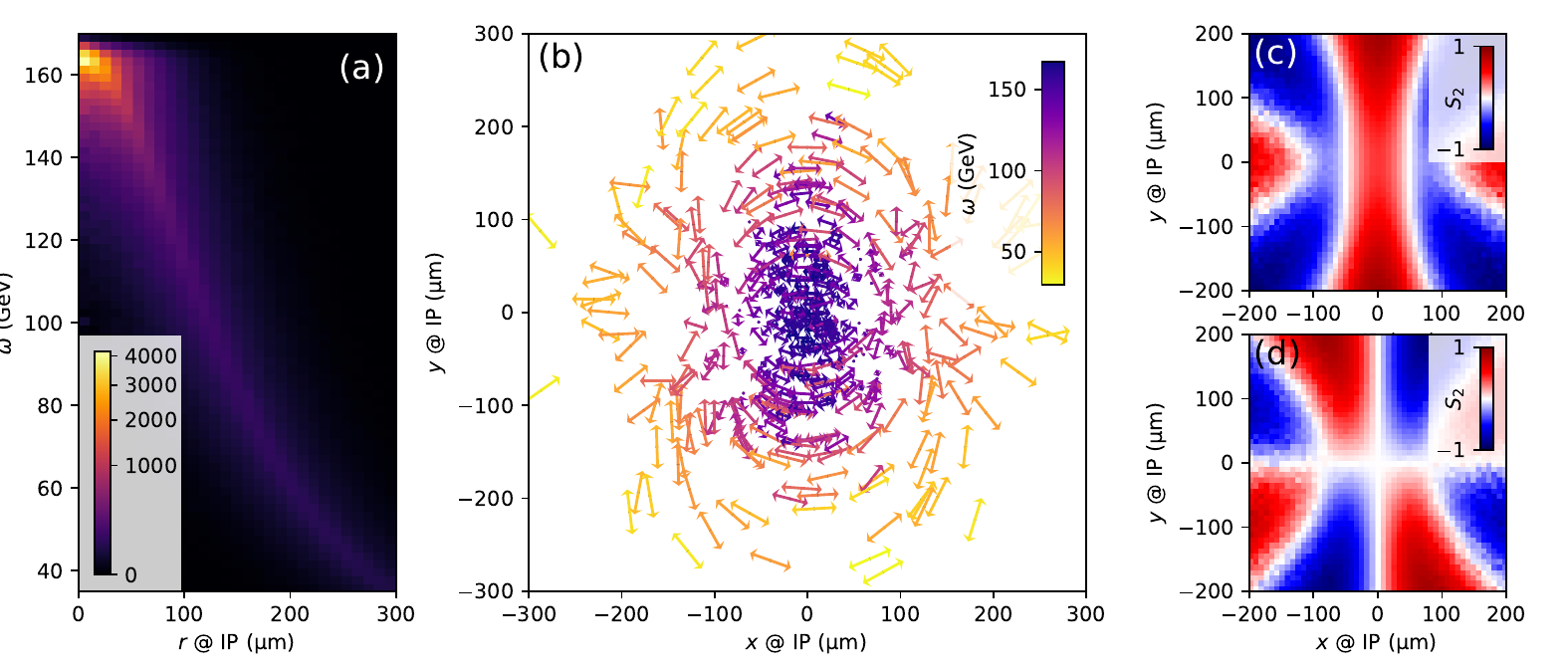}
        \hspace*{-0.6cm}\includegraphics[width=0.96\linewidth]{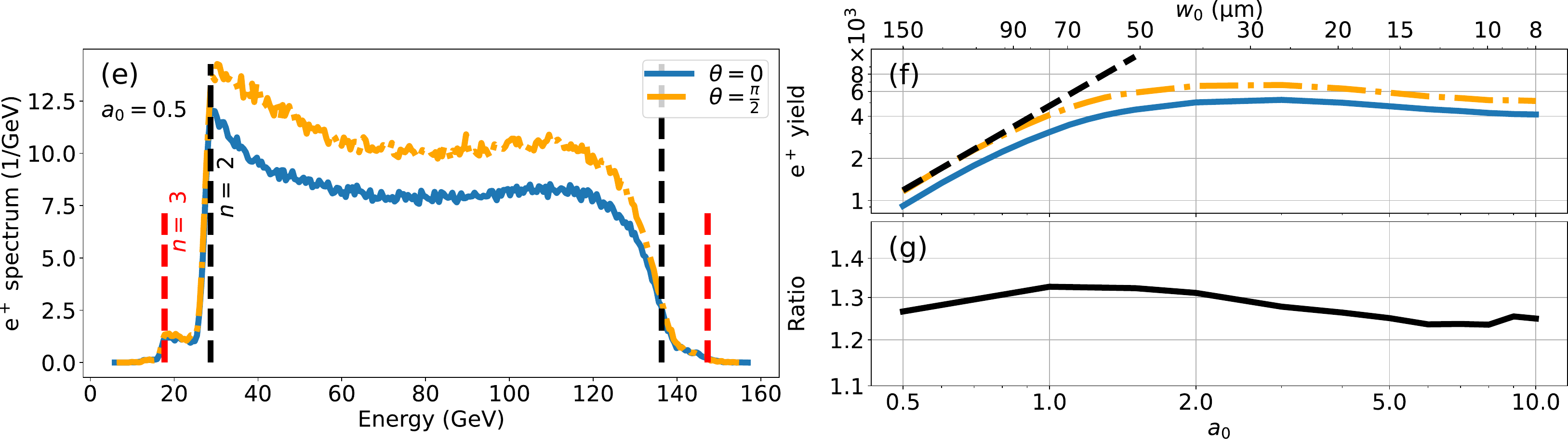}
        \caption{%
            Properties of the $\gamma$-ray beam at the strong-field IP (a-d) and of the positrons they produce (e-g) for case 3 (ILC-like).
            (a) The radially resolved $\gamma$-ray energy spectrum.
            (b) Orientation and magnitude of the polarization (arrow orientation and length), as well as the energy (colour), of a subset of photons at the IP.
            (c) and (d) Stokes parameters $S_1$ and $S_2$, respectively.
            (e) Positron spectrum for $a_0=0.5$ and two pitch angles $\theta =0,\, \pi/2$: vertical dashed lines give harmonic boundaries.
            (f) Total positron yield as a function of $a_0$ (and laser spot size $w_0$) for the two pitch angles; quadratic scaling (black, dashed line).
            (g) The ratio of the yields at $\theta = \pi/2$ and $\theta =0$.
        }
        \label{fig:case3}
    \end{figure*}
    
Finally, we consider what would happen if this experiment were realised at a future linear collider like the ILC \cite{ilc}. With an electron beam of energy 200 GeV and a high-power laser, this would represent a ``Super-LUXE'' experiment with a vastly increased centre-of-mass energy for the SFQED collisions.

For the electron beam emittance, we consider very asymmetric values, in correspondence with the flat beam focusing that is envisioned for future colliders.
The mean photon energy in the IP focal plane is $158$ GeV with an rms spread of $9.2$ GeV, and a mean fluence of 311 photons/µm$^2$ per BX.
The mean polarization degree is 40~\% for phtons in the central 4-µm spot, which decreases to 22~\% and 12~\% when averaging over 100-µm and 300-µm spots, respectively. These values are much smaller than in cases 1 or 2, as indeed in the beam center. The main reason for this behaviour lies in the fact that in case 3 the quantum energy parameter $\eta_\mathrm{ICS}\simeq 2.37$ which means that recoil effects and spin-flips during the Compton scattering become important. This is known to result in non-dipole contributions to the scattered $\gamma$ rays.

In \cref{fig:case3}(e) the positron energy spectrum is plotted for $a_0 =0.5$, where we see an asymmetry due to the asymmetric photon energy distribution. However, there are now two different harmonic orders visible, $n=2$ and $n=3$ ($n=1$ is not present as it is forbidden). The corresponding energy intervals are indicated with vertical dashed lines, assuming that the photons Compton edge is around 165 GeV.

Again we show the number of positrons as a function of $a_0$ (and spot size $w_0$) in \cref{fig:case3}(f). At $a_0 < 1$ the scaling of the yield is approximately quadratic, as indicated by the black dashed line. This may be explained by considering the how the pair probability scales with $a_0$. The number of positrons $N_+ \propto a_0 ^{2n_\star}w_0^2$ where $n_\star$ is the lowest allowed harmonic order and $w_0^2$ is the area of the overlap between the beams. In this case $n_\star = 2$ and $w_0\propto a_0 ^{-1}$ (because the laser energy is fixed ~\cref{eq:spotsize}), so we expect $N_+ \propto a_0 ^2$. At larger $a_0$ a turning over of the positron yield can be observed due to nonlinear effects and the shrinking laser spot size. From \cref{fig:case3}(g) we see that the ratio between the yields is closer to unity, indicating that the pitch angle plays a smaller role for this case due to the increased $\eta$.

\begin{figure}[!ht]
        \centering
        \includegraphics[width=0.8\linewidth]{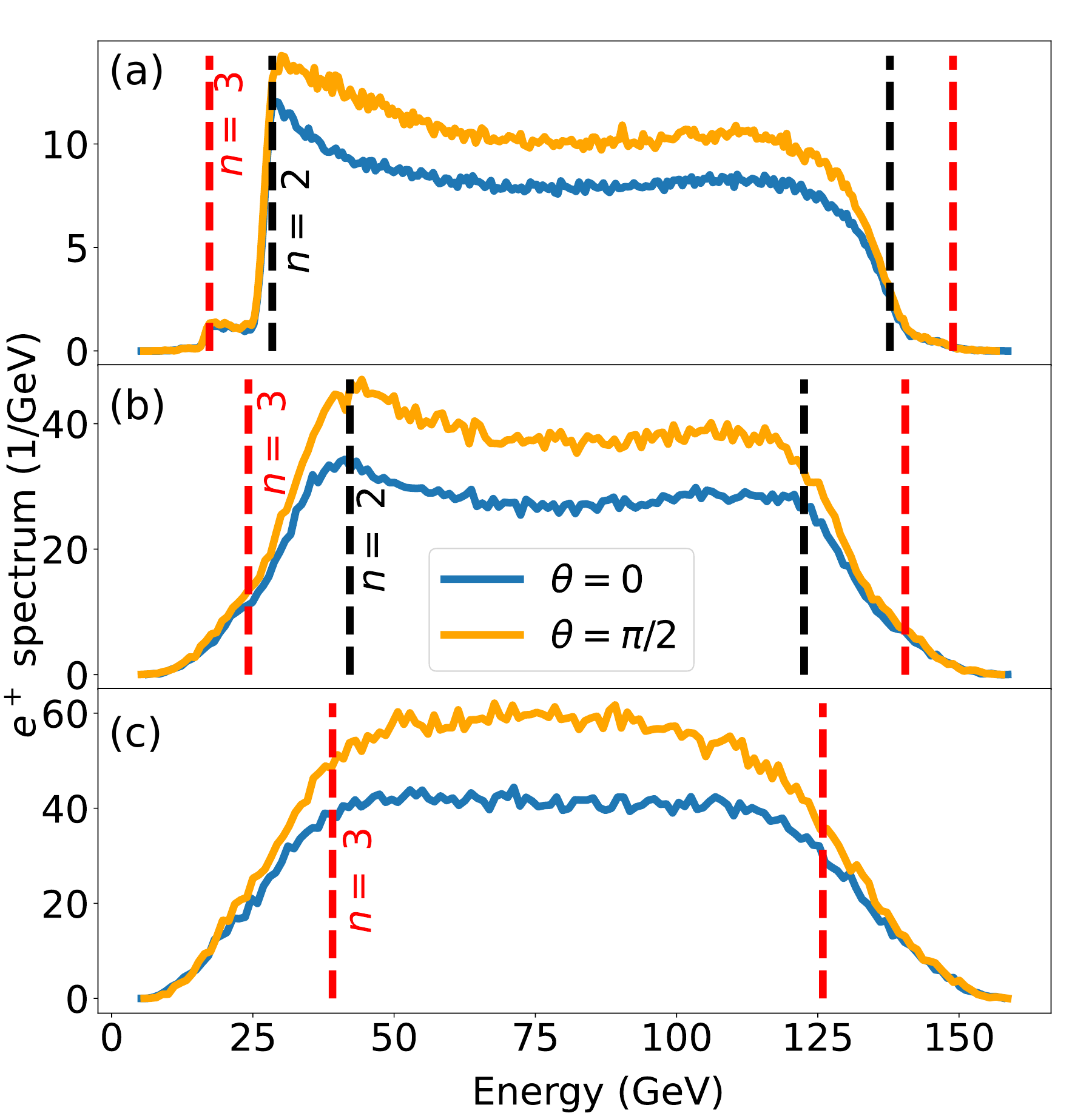}
        \caption{The positron spectrum for (a) $a_0 = 0.5$, (b) $a_0 = 1.0$ and (c) $a_0 =1.5$ for case 3.
        Two pitch angles $\theta = 0$ and $\pi/2$ are considered.
        Vertical dashed lines give the boundaries for the second and third harmonics.}
        \label{fig:harmonicchannels}
    \end{figure}

An additional opportunity in this case, where $\eta > 1$, is to study channel closings~\cite{nousch.plb.2012,titov.epjd.2020,eckey.pra.2022}. 
In \cref{fig:harmonicchannels} we show the positron spectrum for the transition between $a_0 = 0.5$ and $a_0 = 1.5$. As $a_0$ increases, the energy intervals for each harmonic close in towards the midpoint $s=1/2$. Using \cref{eq:Espan} one finds that $n=2$ vanishes for $a_0\approx 1.4$, which may be seen in the way the shape changes in the transition from $a_0=1.0$ to $a_0=1.5$. 
    
\section{Summary}

In this work we have considered the prospects for precision measurements of nonlinear Breit-Wheeler pair production (NLBW), using the collision of inverse Compton scattered (ICS) $\gamma$ rays and a high-intensity laser pulse.
Crucially, the high degree of linear polarization provided by ICS enables the polarization dependence of NLBW to be studied in detail.
The positron yields achievable with the ICS-laser configuration of the LUXE experiment are comparable to those achievable with the bremsstrahlung source, which is already planned.
The 70\% increase in the positron yield when the laser polarization is rotated by 90$\degree$ is large enough to be distinguishable from shot-to-shot fluctuations, given the expected runtime~\cite{LUXE-TDR}.
Increasing the electron-beam energy, and so the centre-of-mass energy, to higher values, makes harmonic structure visible in the positron energy spectra.
A strong-field QED experiment colocated with a future linear collider would also be able to explore the transition from linear to multiphoton pair production and to observe channel closings as a function of the intensity-dependent mass shift.

\section*{Acknowledgements}

The authors acknowledge helpful discussions with members of the LUXE collaboration.
Simulations of the pair production stage were enabled by resources provided by the National Academic Infrastructure for Supercomputing in Sweden (NAISS), partially funded by the Swedish Research Council through grant agreement no. 2022-06725.

\bibliography{references}

\end{document}